%% file: specanalysis.tex
\documentclass[useAMS,usenatbib]{bnmn2e}
\pdfoutput=1
\usepackage{amsmath}
\usepackage{microtype}
\usepackage{mathptmx}
\usepackage{graphicx}
\usepackage{booktabs}
\usepackage{astrojournals}
\usepackage[colorlinks=true,citecolor=blue]{hyperref}

\newcommand{\unit}[1]{\mathrm{#1}}

\input{versioninfo.tex}

\volume{arXiv:0912.2317}
\title{Fitting and Comparison of Models of Radio Spectra}

\author[B. Nikolic]{B. Nikolic\\
  Astrophysics Group, Cavendish Laboratory, Cambridge CB3 0HE, UK
  \\\url{email:b.nikolic@mrao.cam.ac.uk}
  \\\url{http://www.mrao.cam.ac.uk/~bn204/}}

\date{r\bzrrevno (date: \bzrdate)}

\pagerange{\pageref{firstpage}--\pageref{lastpage}; } \pubyear{2009}

\begin{document}

\label{firstpage}
\maketitle

\begin{abstract}
  I describe an approach to fitting and comparison of radio spectra
  based on Bayesian analysis and realised using a new implementation
  of the nested sampling algorithm. Such an approach improves on the
  commonly used maximum-likelihood fitting of radio spectra by
  allowing objective model selection, calculation of the full
  probability distributions of the model parameters and provides a
  natural mechanism for including information other than the measured
  spectra through \emph{priors}.  In this paper I cover the
  theoretical background, the algorithms used and the implementation
  details of the computer code. I also briefly illustrate the method
  with some previously published data for three near-by galaxies. In
  forthcoming papers we will present the results of applying this
  analysis larger data sets, including some new observations, and the
  physical conclusions that can be made.  The computer code as well as
  the overall approach described here may also be useful for analysis
  of other multi-chromatic broad-band observations and possibly also
  photometric redshift estimation. All of the code is publicly
  available, licensed under the GNU General Public License, at
  \url{http://www.mrao.cam.ac.uk/~bn204/galevol/speca/index.html}.
\end{abstract}

\section{Introduction}

The spectrum of radio emission from the majority of astronomical
sources consists of a smooth continuum with (in some but not all
sources) atomic and molecular lines superimposed, most notably the
H{\sc\,i} line at 1.42\,GHz and the carbon monoxide rotational line
ladder starting at 115\,GHz.  Measurements with significant fractional
bandwidths are almost always dominated by the continuum emission.

The shape of the continuum emission can be measured by making
observations at widely spaced frequencies. The available measurements
for some local objects span a very wide range of frequencies, from
about 50\,MHz to above 1\,THz and into the far-infrared portion of the
spectrum. The measured shape of the continuum emission naturally
provides information about the physical properties of the sources and
mechanisms for emission.  There are many examples of physical
properties which one can try to extract:
\begin{itemize}
  \item The slope of the synchrotron emission gives the energy spectrum
    of relativistic electrons in the source, providing constraints on
    the mechanism by which they are created
  \item The change in slope in the synchrotron spectrum gives an
    estimate of the age of the source
  \item The low frequency turn-over, i.e., where source becomes opaque
    due to absorption due to electrons, places constraints on the
    geometry of the source
  \item The slope of the Raleigh-Jeans part of the dust spectrum
    places constraints on the properties of the dust 
  \item The frequency of the peak of the dust spectrum determines the
    temperature of the dust and the total power output of the source
\end{itemize}
Additionally one can try to make an estimate of redshift of a distant
object from the observed spectral shape.

Most of the physics underlying the emission processes at these
wavelengths is well understood and it is possible to calculate the
expected spectrum given a model and its parameters. Therefore,
analysis of radio continuum spectra often consists of `fitting' a
selection of known models to the observations.

Some of the desirable outputs of such an analysis are:
\begin{enumerate}
  \item An objective measure of how well the model explains the data
  \item Unbiased estimates of parameters of the model, estimates of
    errors with which these estimates are made, and the correlations
    between the estimate errors
  \item Full probability distributions of parameters in cases that
    they are significantly non-Gaussian
  \item An objective way of comparing how well different models
    explain the same data
\end{enumerate}

All of these can be obtained simultaneously using Bayesian analysis
and the nested sampling algorithm. In this paper I describe a computer
code to perform this analysis and illustrate it with examples using
previously published data for three near-by galaxies.

This code was developed to support an ongoing observational programme
to measure the radio spectral energy distributions of near-by
star-forming galaxies.  We are also currently applying he code to
analyse the spectra of Ultra-Luminous InfraRed Galaxies (ULIRGs) with
the goal of better understanding the their radio emission and the
physical conditions within them. These results will be published
separately in forthcoming papers.

\section{Method}

The analysis proceeds in the usual fashion, starting with the Bayes
equation \citep[see for example:][]{Jaynes:PTLS,SiviaD06}:
\begin{equation}
  p(\theta | D, H) = \frac{ p(D | \theta, H) p(\theta|H)}{p(D|H)}=
  \frac{ L(\theta)  \pi(\theta)}{Z}
  \label{eq:BayesTheorem}
\end{equation}
where the symbols have following meaning:
\begin{description}

  \item[$H$] is the hypothesis under which the data are analysed. In
    this case, the hypothesis consists of the model we assume for the
    radio emission and the priors for each of the model parameters

  \item[$\theta$] is a vector with elements that are the parameters of
    the model (e.g., the spectral index $\alpha$, the frequency of the
    spectral break $\nu_{\rm br}$)

  \item[$D$] are the observed data (in this case, the observed flux
    densities at various frequencies)

  \item[$p(\theta|H) = \pi(\theta) $] is the probability that the
    model parameters take a particular value, i.e., the prior
    information associated with the hypothesis that we are using

  \item[$p(D|\theta,H)=L(\theta)$] is the likelihood, i.e., the probability of
    observing the data we have given some model parameters $\theta$

  \item[$p(D|H)=Z$] is the so-called Bayesian \emph{evidence}, that is
    a measure of how well our hypothesis (i.e., the model for the
    emission spectrum and the priors) predict the data actually
    observed

  \item[$p(\theta | D,H)$] is the posterior joint distribution of the
    model parameters
\end{description}

The two \emph{inputs} to the calculation are:
\begin{enumerate}
\item $H$, that is a model for the emission spectrum and the prior on
  its parameters (see \ref{sec:synchrotron-models})
\item $L(\theta)$, a function which uses observed data, the model and
  the parameters $\theta$ to calculate the likelihood of a predicted
  spectrum (see \ref{sec:likelihood})
\end{enumerate}

The computation is done using the nested sampling algorithm
\citep{Skilling2006} as described in
Section~\ref{sec:nested-sampling}.

The two outputs are:
\begin{enumerate}
  \item $Z$, the evidence for the model and the prior
  \item $p(\theta | D,H)$, the posterior distribution of the model
    parameters
\end{enumerate}

\subsection{Models of synchrotron radiation spectra}
\label{sec:synchrotron-models}

In general, the models that it makes sense to try when analysing a
particular set of observations are determined by the type of object
that has been observed and the region of the spectrum which is being
analysed. The current version of the computer code described here
already has implementations of models which combine absorption,
synchrotron and thermal radiation.

In this paper however, I restrict the description to models of
non-thermal synchrotron emission. The reason is that in the examples
shown later I use only measurements below 5\,GHz while the synchrotron
mechanism is the dominant component of emission from the majority of
extragalactic sources at frequencies below about 50\,GHz, so this is a
reasonable approximation. The remaining models present in the software
implementation will be used for forthcoming science papers and they
can easily be extended further still with thermal dust or spinning
dust emission for example.

The simplest model of synchrotron emission spectrum is a simple
power-law model:
\begin{equation}
  F_{\nu}(\nu)= F_{\nu}^{0} \cdot \left(\frac{\nu}{1\,\unit{GHz}}\right)^\alpha
\end{equation}
with two parameters:
\begin{description}
  \item[$F_{\nu}^{0}$] The flux density at the frequency of 1\,GHz
  \item[$\alpha$] The spectral index
\end{description}
It is not however convenient to make a parametrisation directly in
terms of $F_{\nu}^{0}$ since it can take large range of values for
typical sources, and we do not \emph{a-prior} know even the order
magnitude of $F_{\nu}^{0}$ to expect. For example, if we assumed that
$F_{\nu}^{0}$ could be any value between 0.01 and 1\,Jy with uniform
probability then there would be far more values with
\emph{magnitude\/} of order of 1 than of order 0.01.  Instead, I
parametrise the model in terms of the logarithm of $F_{\nu}^{0}$,
i.e., $\log_{10} (F_{\nu}^{0})$ and assume that this is uniformly
distributed over a range of values (-2 to 0 in the example above).
There is further discussion of this topic of so-called \emph{scale\/}
parameters by \cite{Jaynes:PTLS}.

A more complicated and physically more accurate model is the so called
continuous-injection model \citep[e.g.,][]{1962SvA.....6..317K}. In
this model it is assumed that the energetic electrons have a power-law
energy distribution when they are created, and that these
freshly-created electrons are continuously added to the plasma at a
constant rate. As the electrons emit synchrotron radiation they
naturally loose energy.  The rate of energy losses is however higher
for the higher-energy electrons, and this leads to a `break' in the
emission spectrum of the plasma. The resulting spectrum can be
\emph{approximately} described by:
\begin{align}
  F_{\nu}(\nu)= \left\{
      \begin{array}{lr}
        \displaystyle 
      F_{\nu}^{0} \cdot \left(\frac{\nu}{1\,\unit{GHz}}\right)^\alpha
      &
      \nu \leq \nu_{\rm br}
      \\
      \displaystyle 
      F_{\nu}^{0} \cdot
      \left(\frac{\nu}{1\,\unit{GHz}}\right)^{\alpha}
      \left(\frac{\nu}{\nu_{\rm br}}\right)^{-1/2} 
      &
      \nu > \nu_{\rm br}.
  \end{array}
  \right.
\end{align}
The one new parameter in this model is the frequency of break in the
spectrum, $\nu_{\rm br}$. If this break frequency is known, it can be
used to estimate the age of the source given the magnetic field
strength within it and vice-versa.

Like the flux density scaling parameter $F_{\nu}^{0}$, the frequency
of the spectral break is a scale parameter and is best parametrised as
the logarithm of the actual frequency: $\log_{10}(\nu_{\rm br})$.

The models just described here are continuous functions of frequency
while actual measurements are integrated over some finite
bandwidth. The effect of averaging over a bandwidth can be taken into
account analytically for the simple models described here or more
generally by numerically integrating the average flux over the
band. In this work however, I however make the approximation that the
bandwidths are small and that it can be assumed without too great an
error that each measurement is monochromatic. 

The above two models describe the intrinsic emission spectrum from
relativistic electrons. Depending on the geometry of the source and
total density of electrons, the electrons may absorb as well as emit
radiation. This self-absorption process usually becomes significant at
a low enough frequencies, leading to a turn-over in the spectrum. The
self absorption factor, $A_{\rm s}$
\citep[e.g.,][]{1970ranp.book.....P} can be parametrised by the
frequency at which emission peaks, $\nu_{\rm pk}$:
\begin{align}
  x&= \nu/\nu_{\rm pk}\\
  A_{\rm s}&= x^{-\alpha+5/2} \left[1- \exp\left(1-x^{\alpha-5/2}\right)\right]
\end{align}
Again, this quantity is an additional parameter of models and should
be parametrised in terms of its logarithm: $\log_{10}( \nu_{\rm pk})$.

The possibility of Synchrotron-Self Absorption (SSA) leads to a total
of four possible models in this simple example: power-law, power-law
that also exhibits self-absorption, continuous-injection and
continuous-injection with self-absorption.

As described above, these models need to be combined with priors on
their parameters for them to be useful. For this illustration I use
simple priors with following two properties which ease analysis:
\begin{enumerate}
\item The prior factors into independent functions of one parameter
  only; e.g., for the most complex model:
  \begin{multline}
    \pi(\{\log_{10}\left(F_{\nu}^{0}\right), \alpha,
  \log_{10}(\nu_{\rm br}), \log_{10}(\nu_{\rm
    pk})\})= \\ 
  \pi\left[\log_{10}\left(F_{\nu}^{0}\right)\right]\cdot\pi(\alpha)\cdot
  \pi\left[\log_{10}(\nu_{\rm br})\right]\cdot
  \pi\left[\log_{10}(\nu_{\rm pk})\right]
\end{multline}
\item Each prior is constant within a given range and zero outside:
\begin{equation}
\pi(x)=\begin{cases}
  \frac{1}{x_{\rm high}-x_{\rm low}} & x_{\rm low}<x<x_{\rm high}\\
  0  & \text{otherwise}
\end{cases}
\end{equation}  

For all of the examples presented below I used the same set of priors
as follows:
\begin{align}
  \log_{10}\left(F_{\nu}^{0}\right)_{\rm low}&=-1  &
  \log_{10}\left(F_{\nu}^{0}\right)_{\rm high}&=1 
\label{eq:priors-bg} \\
  \alpha_{\rm low}&=-3/2 &  \alpha_{\rm high}=1/2\\
  \log_{10}(\nu_{\rm br})_{\rm low} &= 8.5&
  \log_{10}(\nu_{\rm br})_{\rm high} &= 10.0\\
  \log_{10}(\nu_{\rm pk})_{\rm low} &=7.5 &
  \log_{10}(\nu_{\rm pk})_{\rm low} &=8.5 .
\label{eq:priors-end}
\end{align}

\end{enumerate}

\subsection{Likelihood}
\label{sec:likelihood}

In this section I describe the calculation of the likelihood of the
observed data given a model and its parameters. This part of the
calculation depends on a good understanding of the observations
and how they were processed so that realistic error estimates on
measured flux densities can be assigned and any selection effects
taken into account. 

In the simplest cases, it is possible to make two simplifying
assumptions:
\begin{enumerate}
   \item Measurements at each frequency are independent, and the
     likelihood therefore factors into a product of functions of flux
     densities at one frequency only
   \item Errors on each measurement are normally distributed, and the
     likelihood therefore takes the standard Gaussian form
\end{enumerate}
If these assumptions are made, than the joint likelihood is simply:
\begin{equation}
  \log{L(\theta)} = 
 - \frac{1}{2}\sum_{i} \left\{
   \left[
   \frac{D_{i}-F_{\nu}(\nu_i)}
   {\sigma_{i}} \right]^{2}
 + \log\left(2\pi \sigma_i^2\right)
 \right\}
\end{equation}
where $D_i$ is the flux density observed at a frequency of $\nu_i$ and
$\sigma_i$ is the estimate of the standard error of this
measurement. The second term in the sum is the normalisation constant
(i.e., it does not depend on the observed data or the model) which
must be included for correct calculation of the evidence value.

The above assumptions are the same as often made in analysis of radio
spectra using more conventional techniques. One of the advantages of
the present approach however is that these assumptions do \emph{not\/}
need to be made.

For example, in multi-frequency surveys of relatively faint sources it
is often the case that sources are selected at the lower frequency
(where the survey speed is typically higher) and only the detected
sources are followed up at the higher frequency. This leads to a
cross-dependence in the likelihood function between the measurements
at the lower and higher frequencies. 

\subsection{Nested sampling}
\label{sec:nested-sampling}

The method I use for the calculation of the evidence, $Z$ and the
posterior distribution of the model parameters, $p(\theta| D, H)$, is
the \emph{nested sampling} algorithm described by
\cite{Skilling2006}. The key advantage of this algorithm for this
application is that it allows efficient and accurate calculation of
the evidence value even in the presence of relatively complex
likelihood distributions.

The starting set of points used by the sampler is initialised by
randomly and uniformly distributing the points in the space allowed by
the priors. Since all priors are flat this is sufficient to ensure a
representative starting distribution.

The sampling then proceeds by finding the point in the current set
with the smallest likelihood and replacing it. The replacement needs
to be selected uniformly from the prior space with the constraint that
the likelihood of the replacement point is greater than the likelihood
of the point it replaces. This is implemented by:
\begin{enumerate}
  \item Selecting a point at random in the current set 
  \item Using a Markov chain with to find a new point subject to the
    likelihood constraint 
\end{enumerate}
The step size and directions used in the Markov chain are determined
by spread of the points in the current set. Specifically, a principal
component analysis is carried out on the current set and the steps in
the Markov chain alternate between each of the eigenvectors, which are
scaled by 0.1 before being used. Normally 100 steps are made with the
Markov chain before the new point is added to the current set. 

The nested sampling procedure is terminated when the requested number
of samples has been made or when the Markov chain procedure fails to
find a point with better likelihood than the worst point in the
set. Because of this latter mechanism for termination of the sampler,
the number of samples made should be inspected before further analysis
to ensure the sampling proceeded far enough to provide accurate
results.

This procedure generally appears to work well but it should be noted
that for multi-modal distributions with widely separated peaks it will
generate step sizes which are too large and with too high a likelihood
of leading to a lower-likelihood region. This would manifest itself as
early termination of the nested sampling algorithm because the Markov
chain constrained sampling does not produce a sample with a higher
likelihood than the worst point in the live set.

I note again that the models I have considered so far \emph{are\/}
(generally) multi-modal, but the modes are sufficiently close that the
present scheme works well. If the models of radio spectra and the
likelihood function are extended further to more complex problems,
this potential problem presented by multi-modal distributions should
be kept in mind.

Significantly more advanced implementations of nested sampling
algorithm are described by \cite{2008MNRAS.384..449F} and
\cite{2009MNRAS.398.1601F}. I believe however that the present
implementation is the only that is publicly available under the GPL
and callable from C++ and Python.

\subsection{Presenting the results}

The evidence values calculated are simple numbers and can be tabulated
for various combinations of models and priors. In the examples below
only one set are priors is used, so only the one evidence value is
given for each model. The relative magnitudes of the evidence allow
objective model selection, with the higher evidence value implying the
preferred model.

For each model, the nested-sampling algorithm also provides the full
joint posterior distribution of the parameters. These are
conventionally visualised by marginalising to get to the marginal
distributions of single parameters and of pairs of parameters.  These
can be then plotted as one-dimensional histograms and two-dimensional
colour plots respectively.

It is clearly also useful to visualise how well each combination of
model and priors fits the observed data. For example, this often
provides crucial information about how the models might need to be
modified to explain the observations. Since the result of the
nested-sampling analysis is a distribution of model parameters, there
are various choices as to how the resulting models may be
visualised. The simplest choice is to plot the model with the
maximum-likelihood parameter set for each model. Good maximum
likelihood estimate can obtained simply by taking the
highest-likelihood point in the live set of the sampler at the
completion of the algorithm. 

\begin{figure*}
  \begin{tabular}{cc}
  Power-law (Evidence: 185.874) & Power-law + SSA (Evidence: 76.3034)\\
  \includegraphics[clip,width=0.49\linewidth]{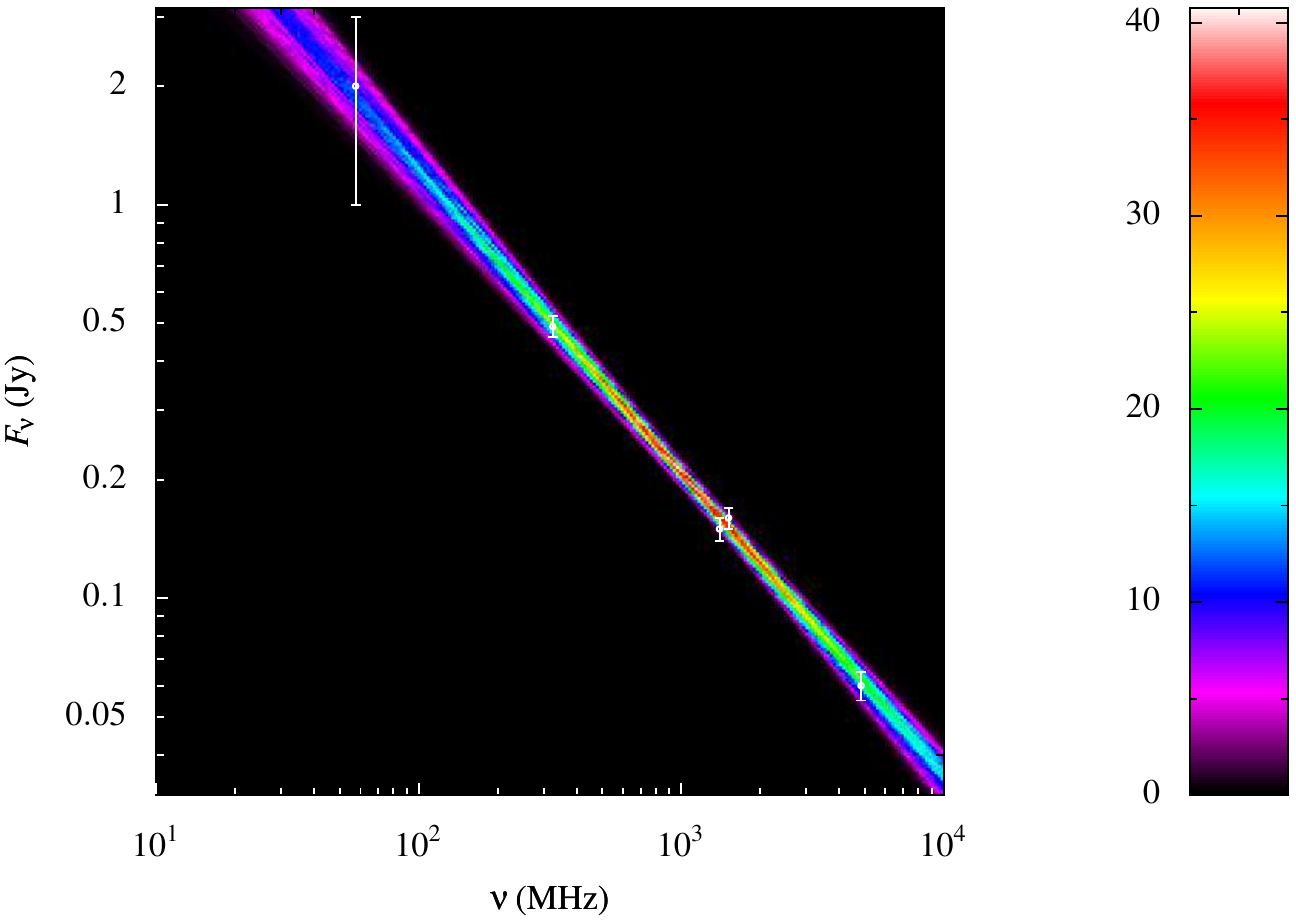}
  &
  \includegraphics[clip,width=0.49\linewidth]{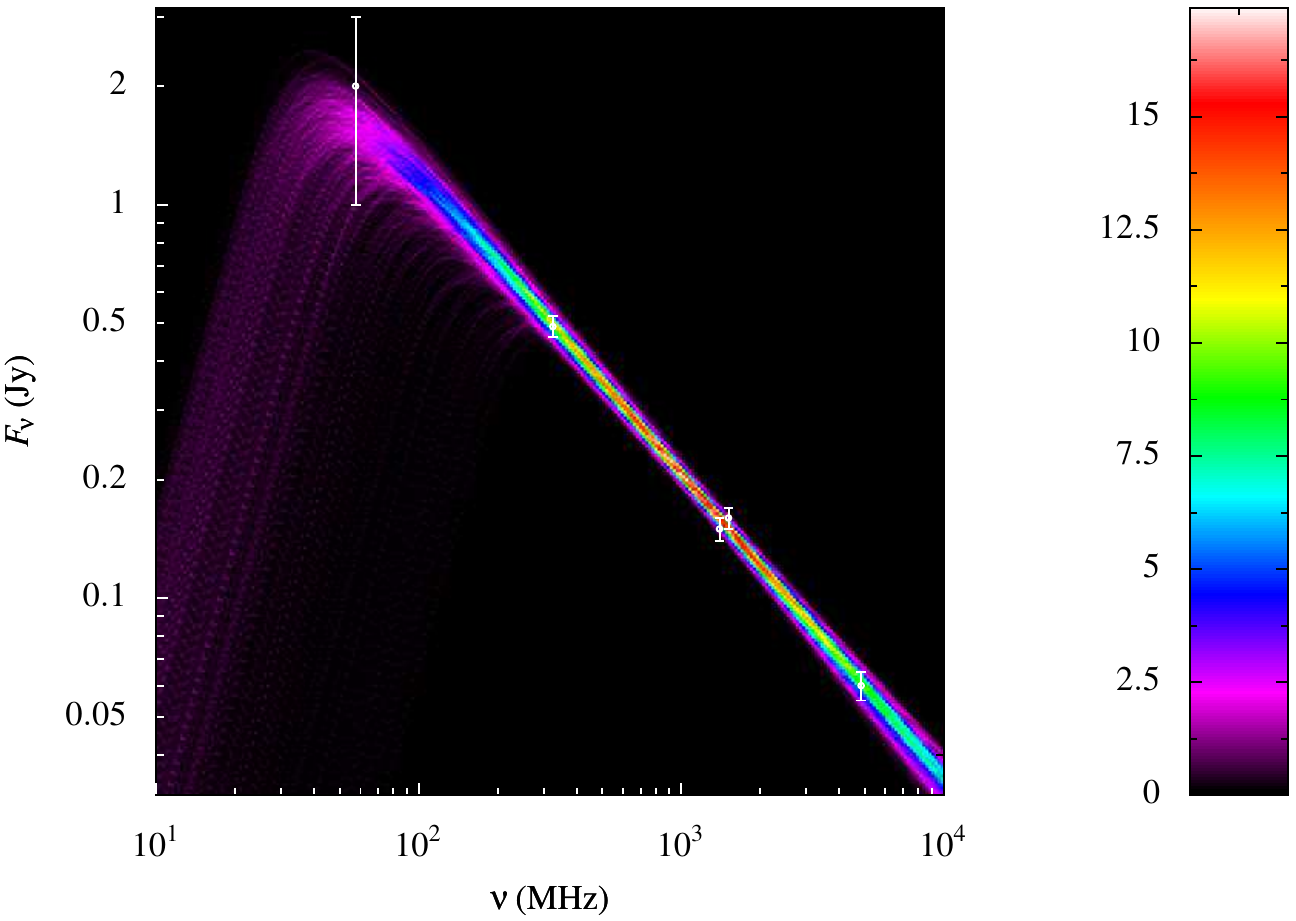}\\
  CI (Evidence: 83.4132) & CI + SSA (Evidence: 27.65)\\
  \includegraphics[clip,width=0.49\linewidth]{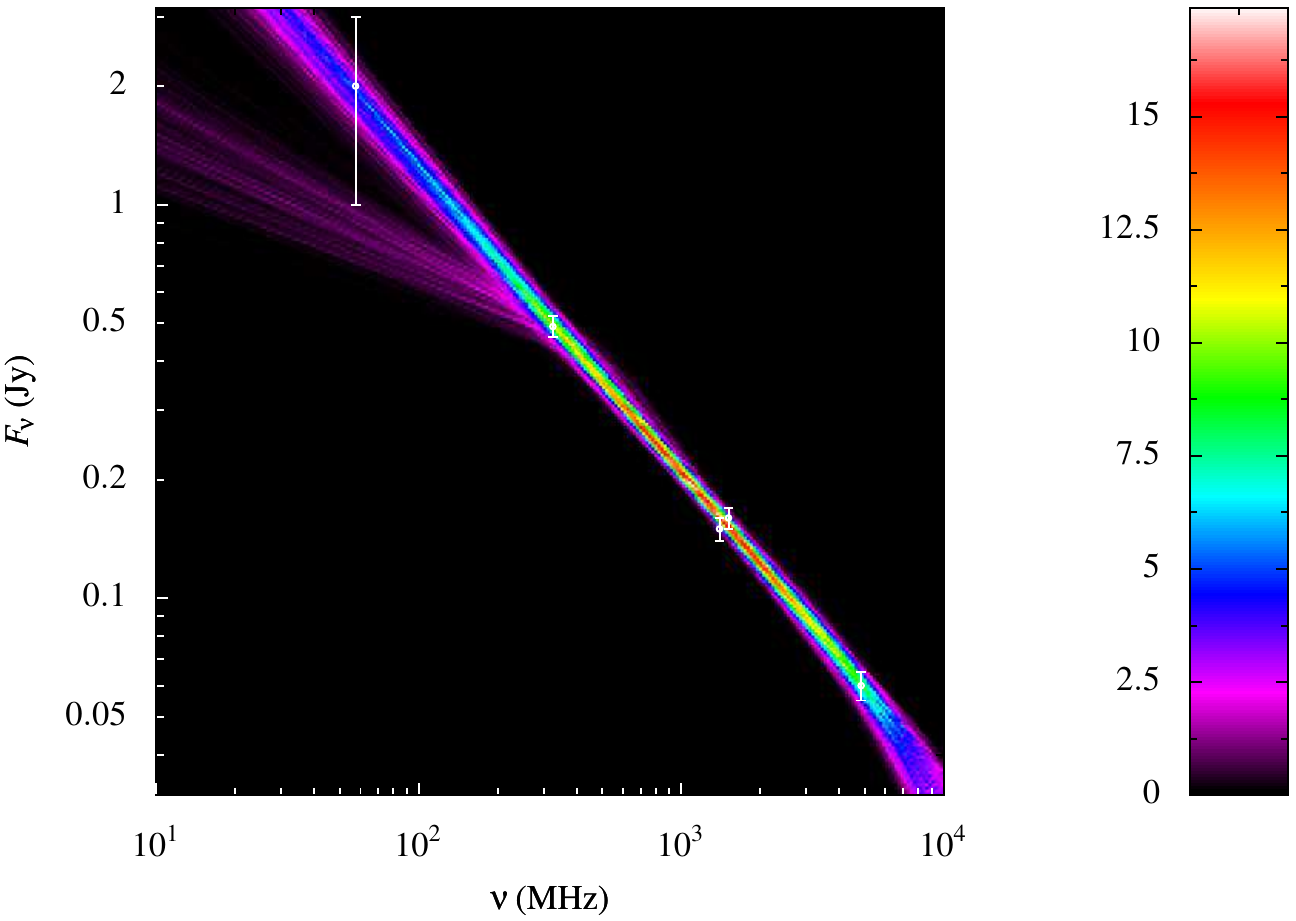}&
  \includegraphics[clip,width=0.49\linewidth]{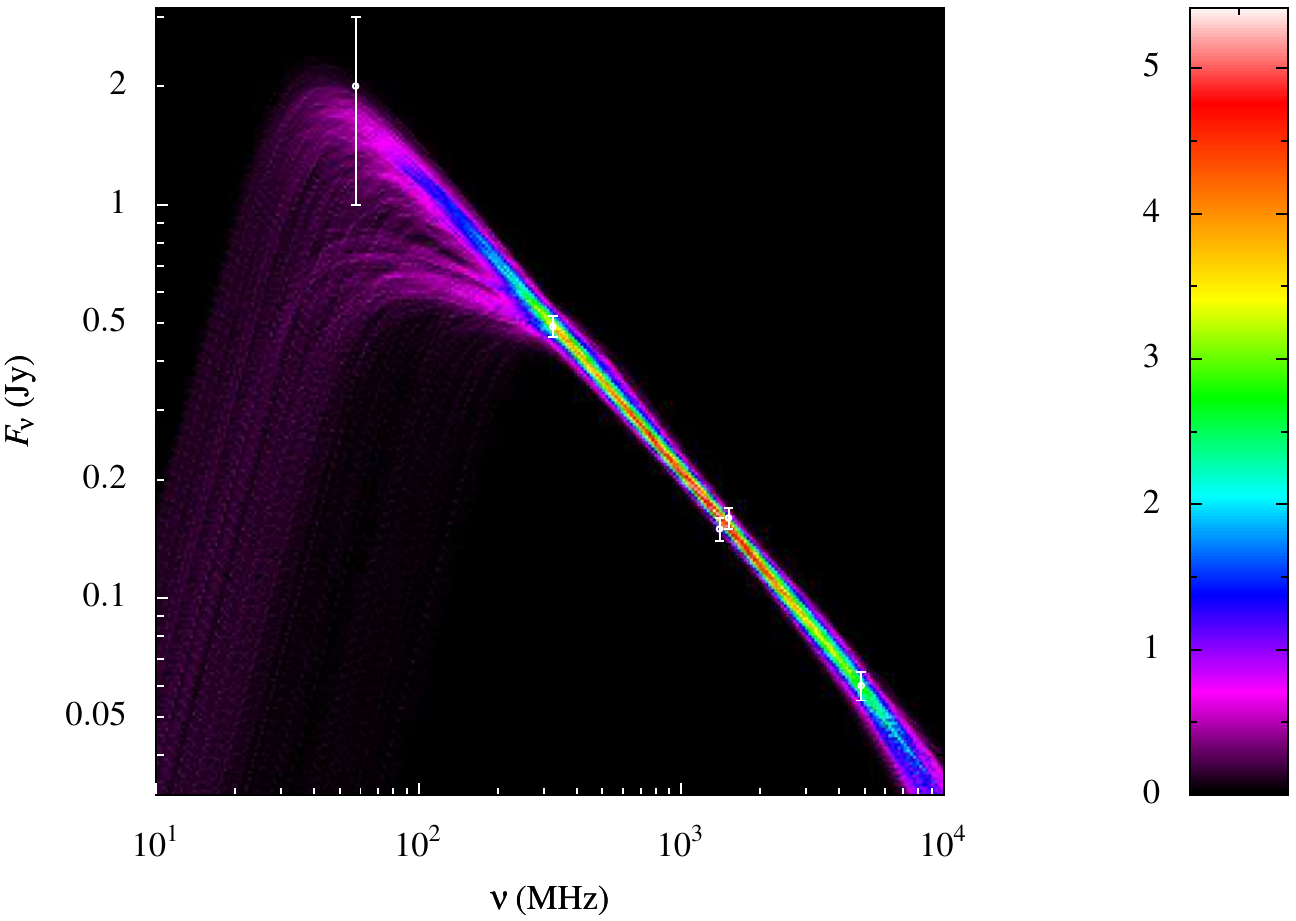}
\end{tabular}
  \caption{NGC 628}
\label{fig:ngc628-fan}
\end{figure*}

\begin{figure*}
\begin{tabular}{cc}
  Power-law (Evidence: $9.7\times 10^{-15}$) & 
  Power-law + SSA (Evidence: $3.9\times10^{-15}$)\\
  \includegraphics[clip,width=0.49\linewidth]{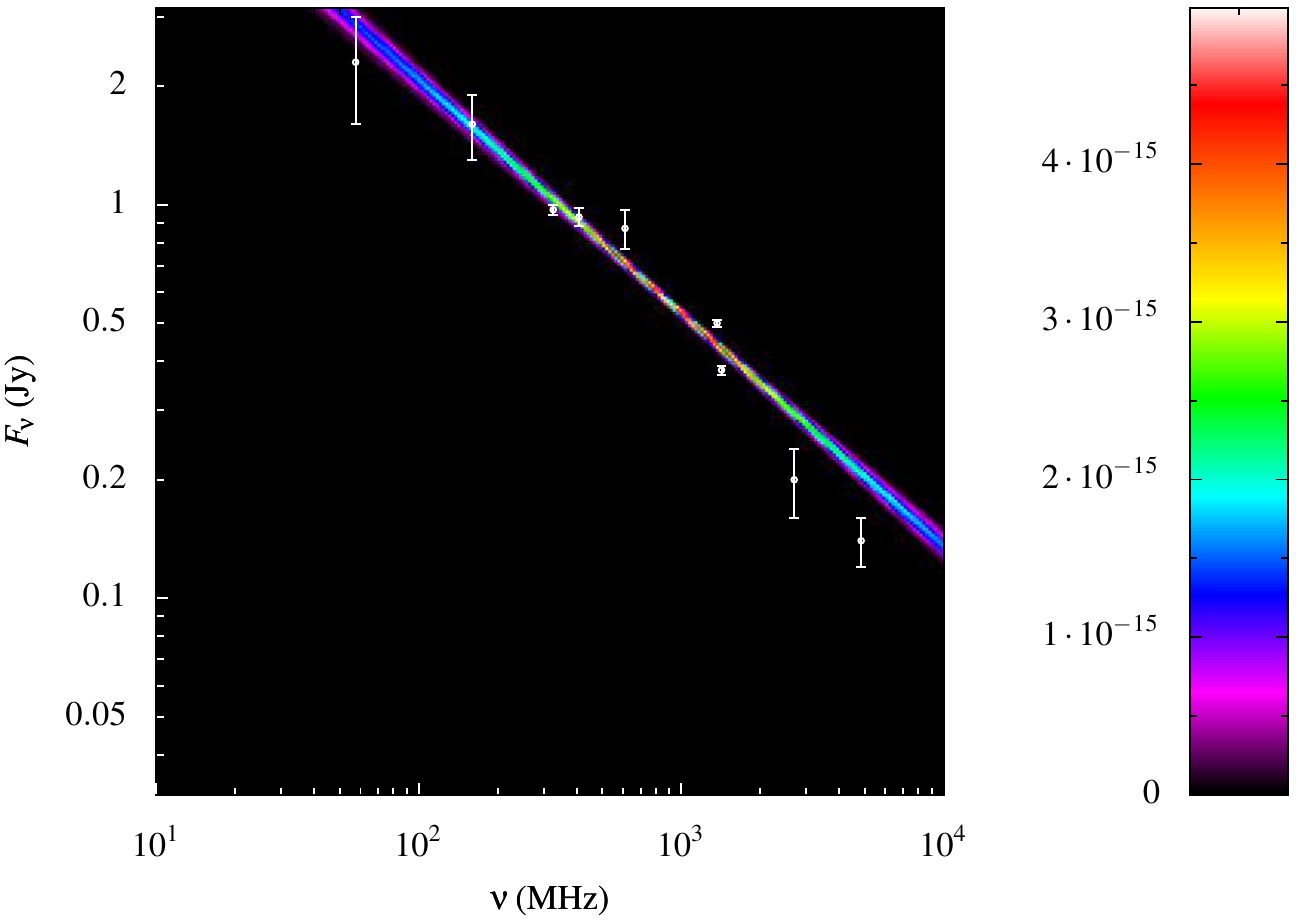}
  &
  \includegraphics[clip,width=0.49\linewidth]{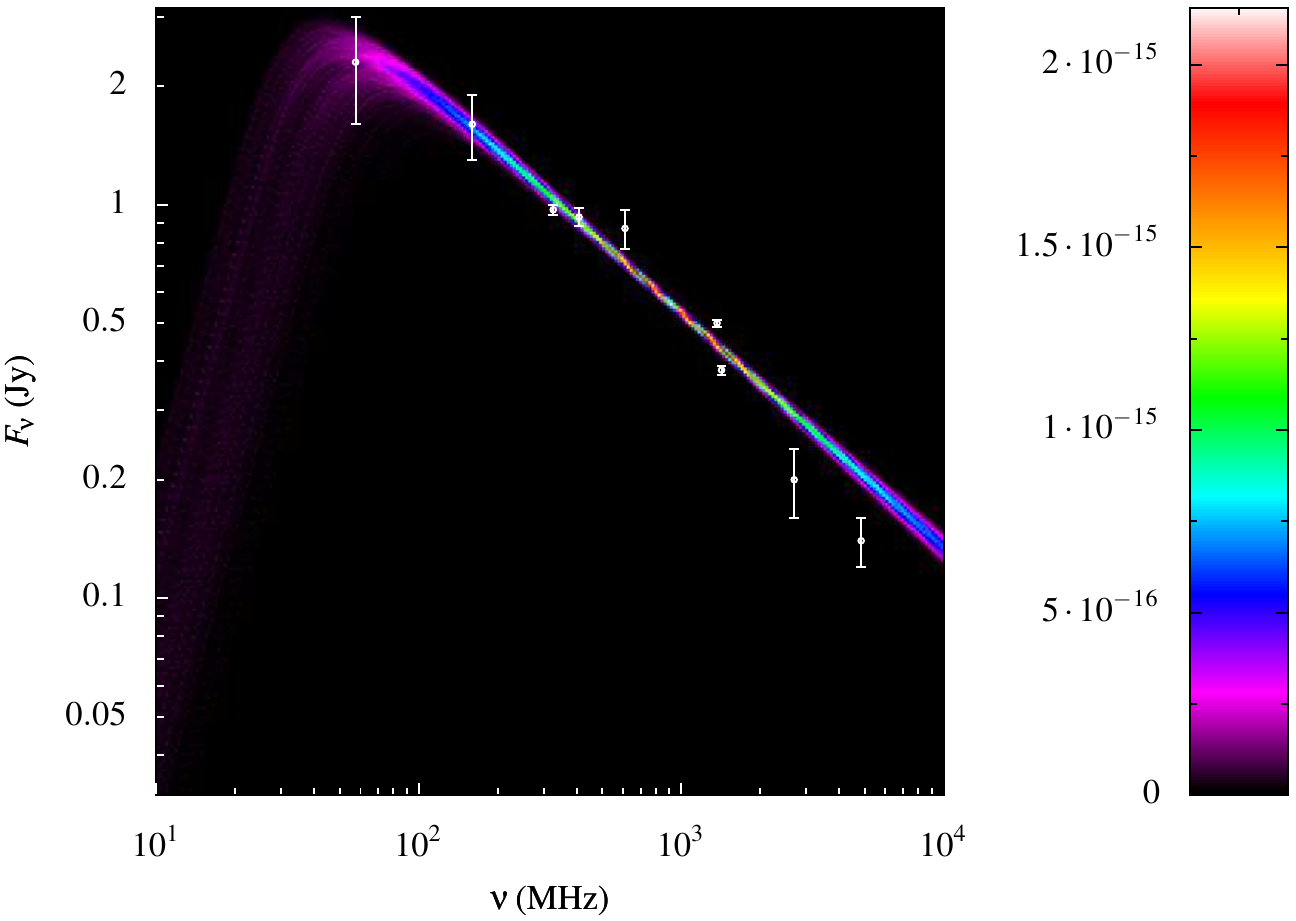}\\
  CI (Evidence: $1.3\times10^{-09}$) & 
  CI + SSA (Evidence: $2.6\times10^{-10}$)\\
  \includegraphics[clip,width=0.49\linewidth]{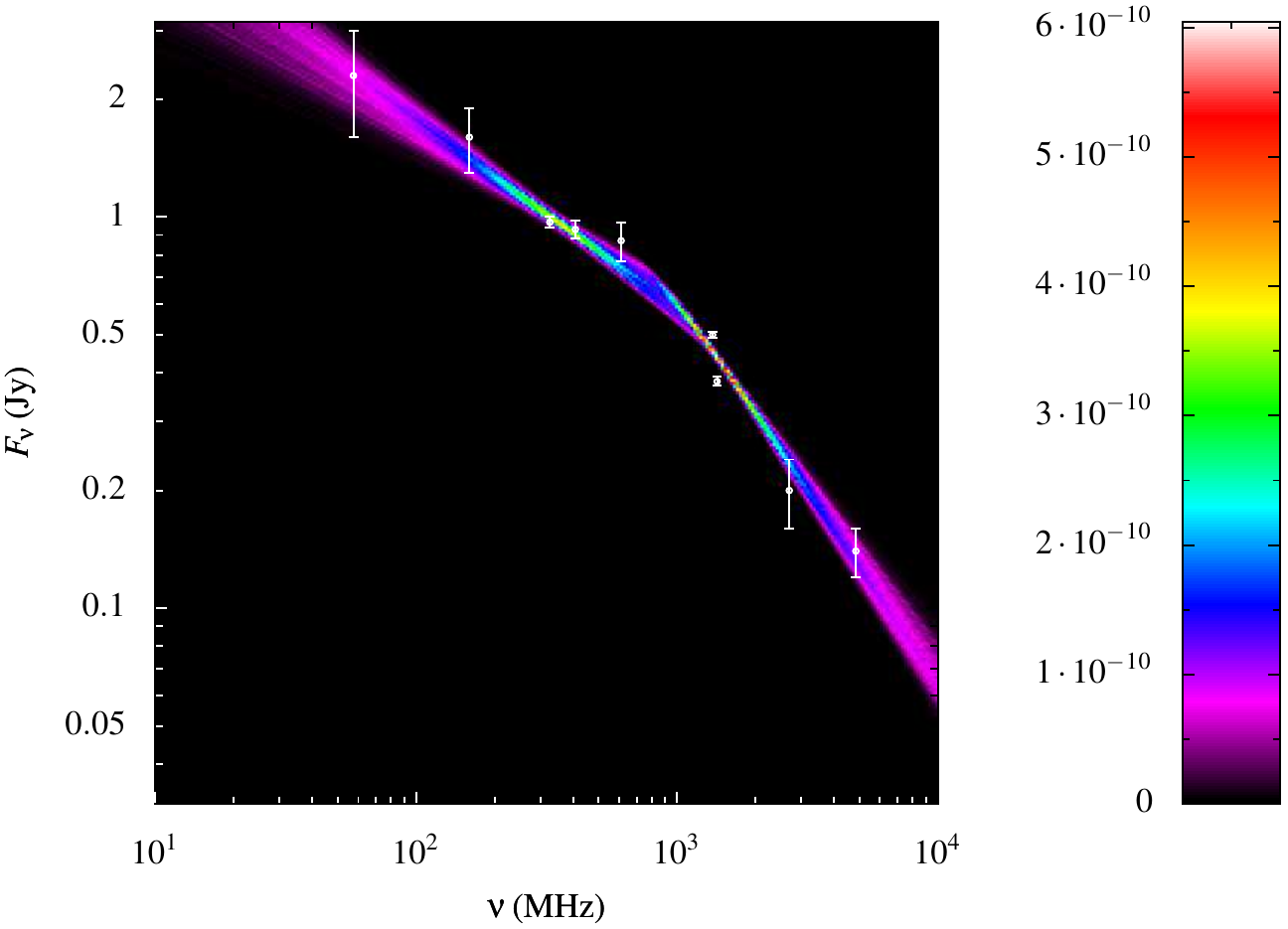}&
  \includegraphics[clip,width=0.49\linewidth]{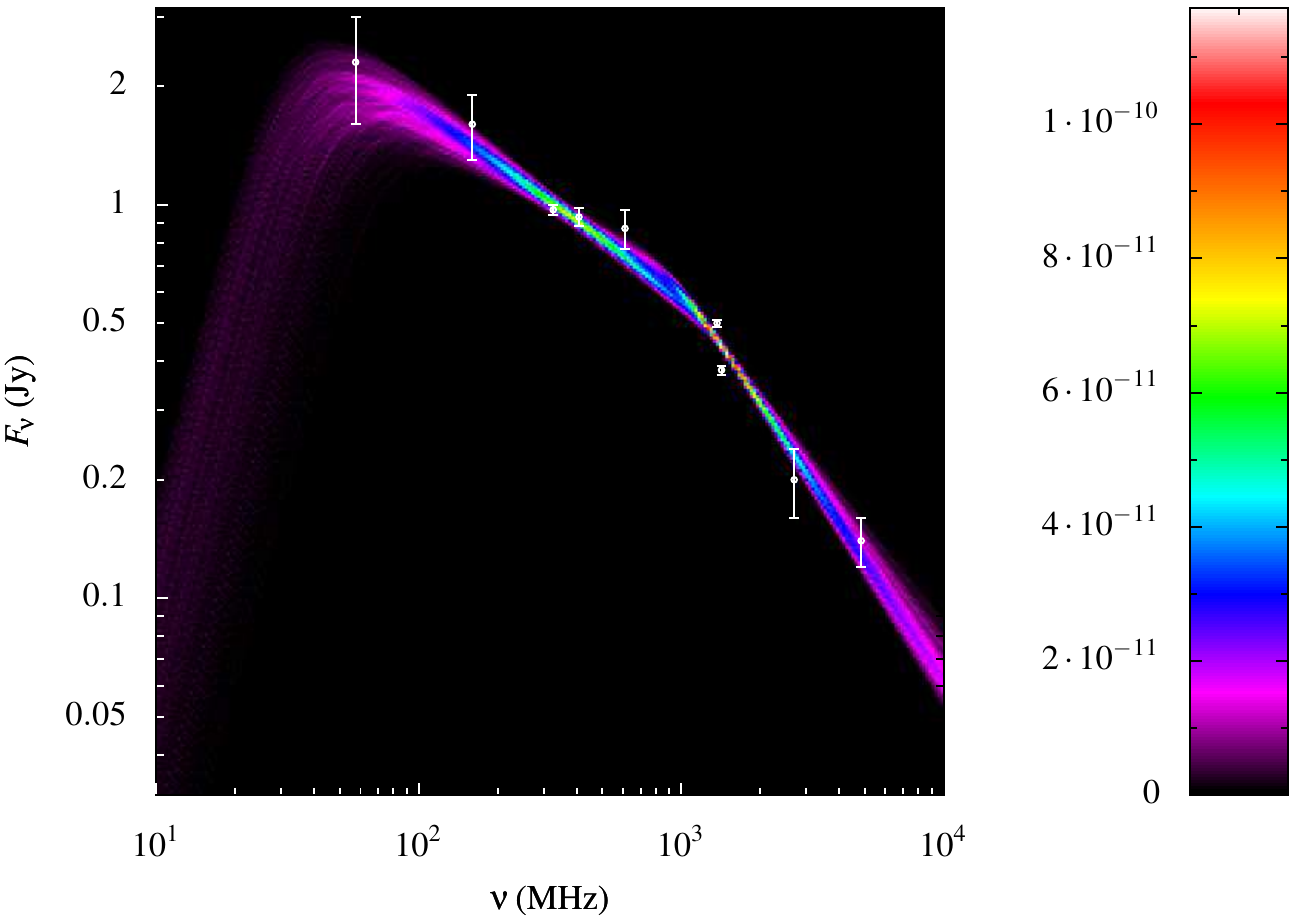}
\end{tabular}
\caption{NGC 3627}
\label{fig:ngc3627-fan}
\end{figure*}

\begin{figure*}
\begin{tabular}{cc}
  Power-law (Evidence: $2.9\times10^{-17}$) & 
  Power-law + SSA (Evidence: 0.5)\\
  \includegraphics[clip,width=0.49\linewidth]{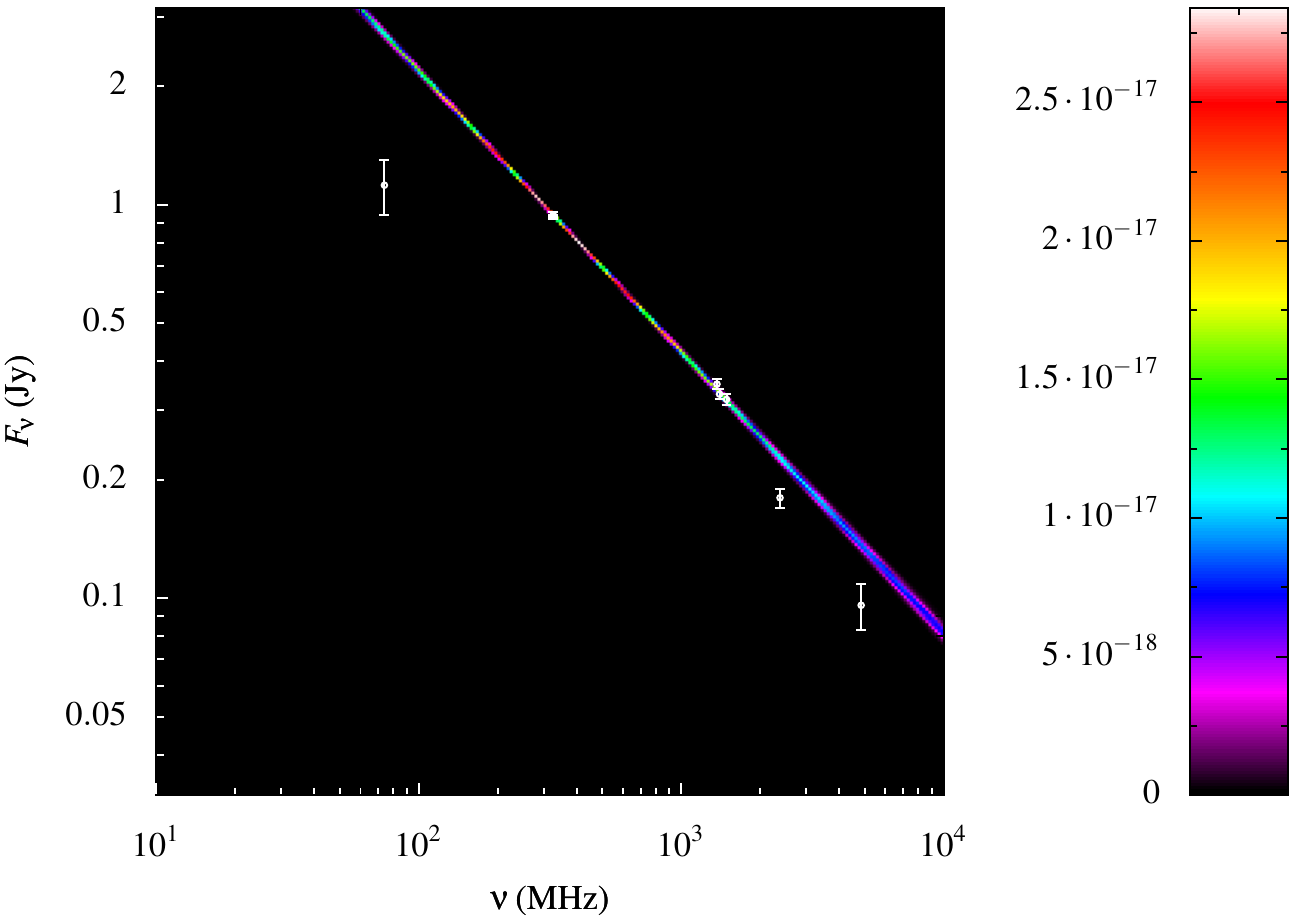}
  &
  \includegraphics[clip,width=0.49\linewidth]{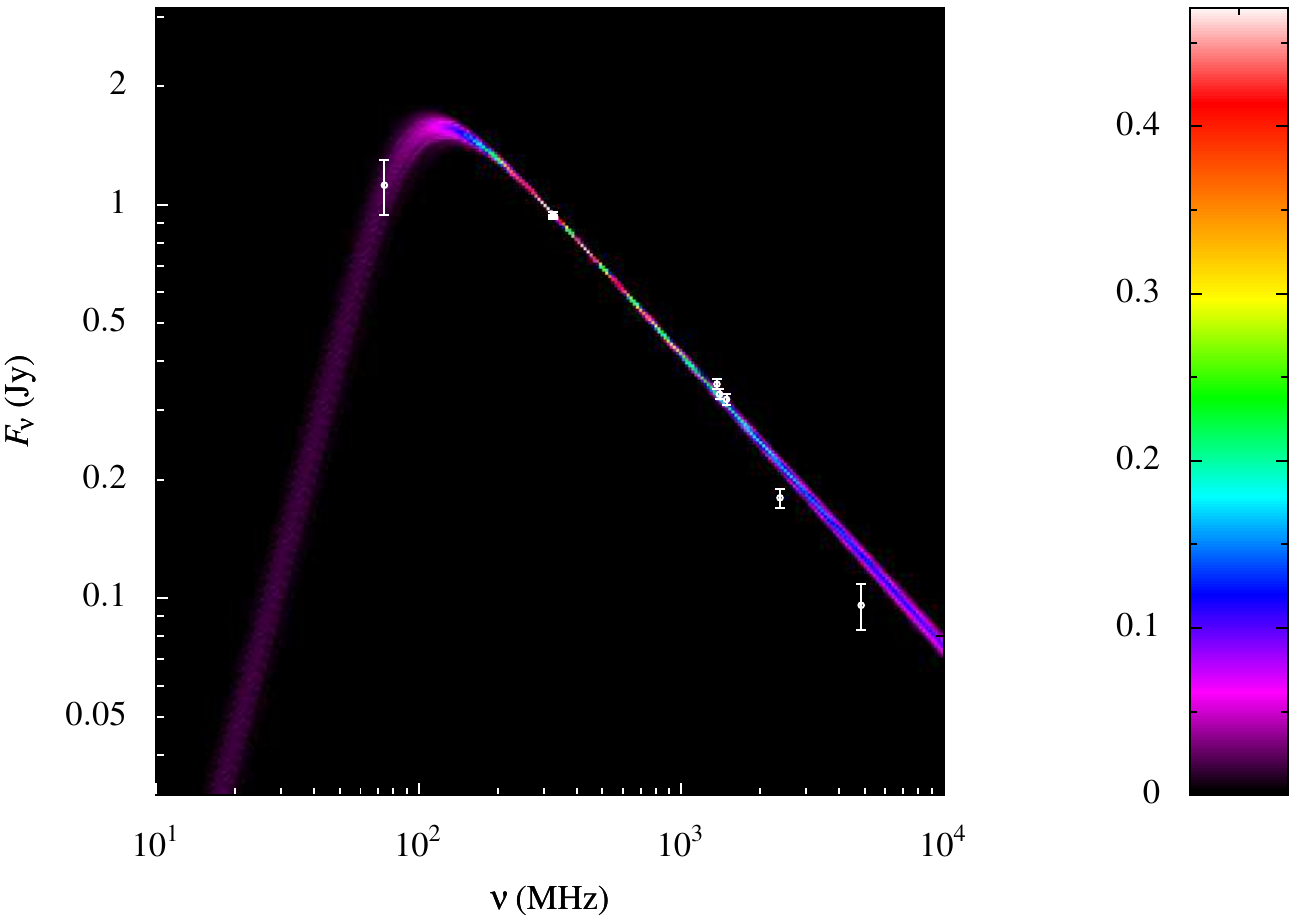}\\
  CI (Evidence: 70) & 
  CI + SSA (Evidence: $5.0\times10^{5}$)\\
  \includegraphics[clip,width=0.49\linewidth]{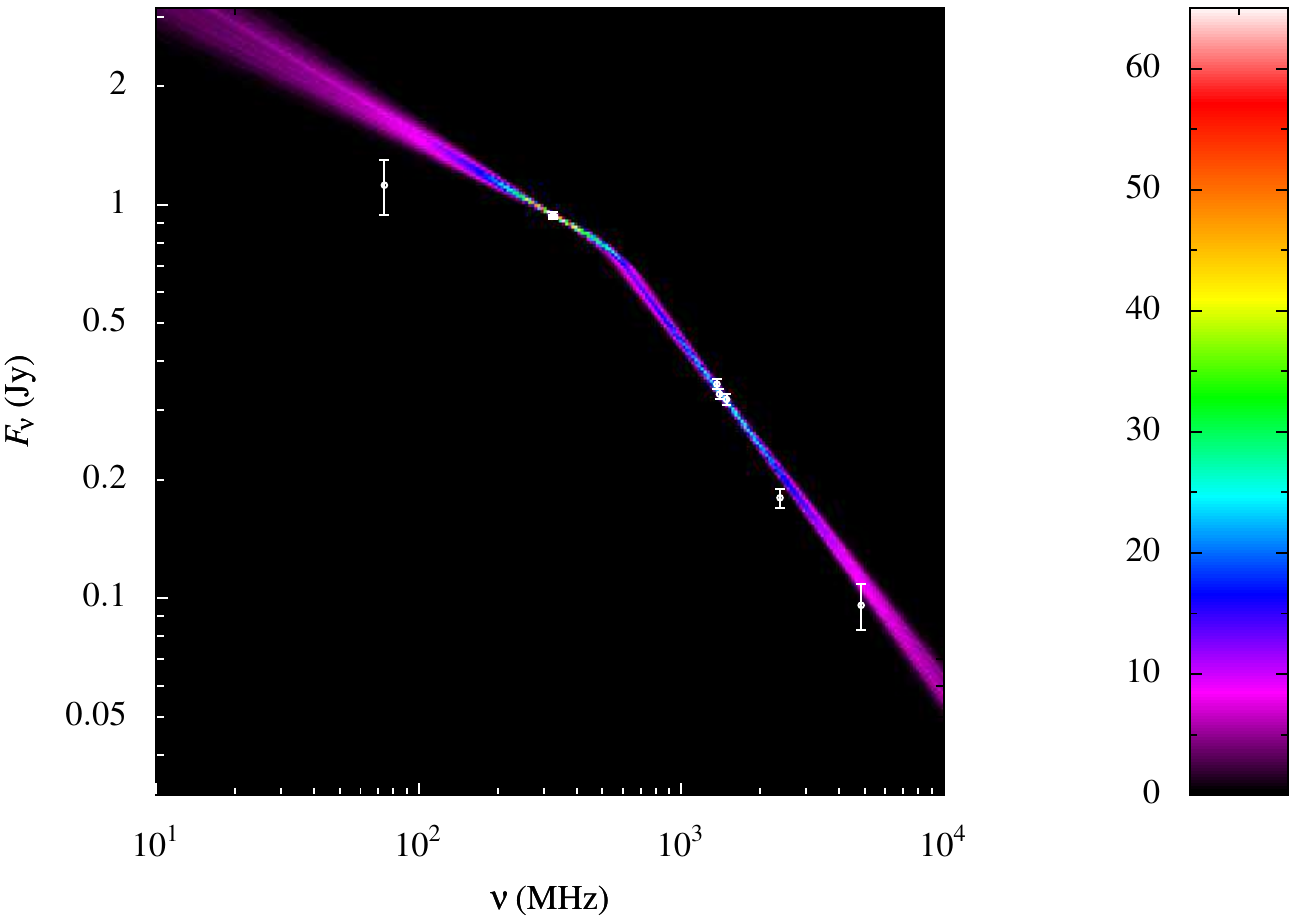}&
  \includegraphics[clip,width=0.49\linewidth]{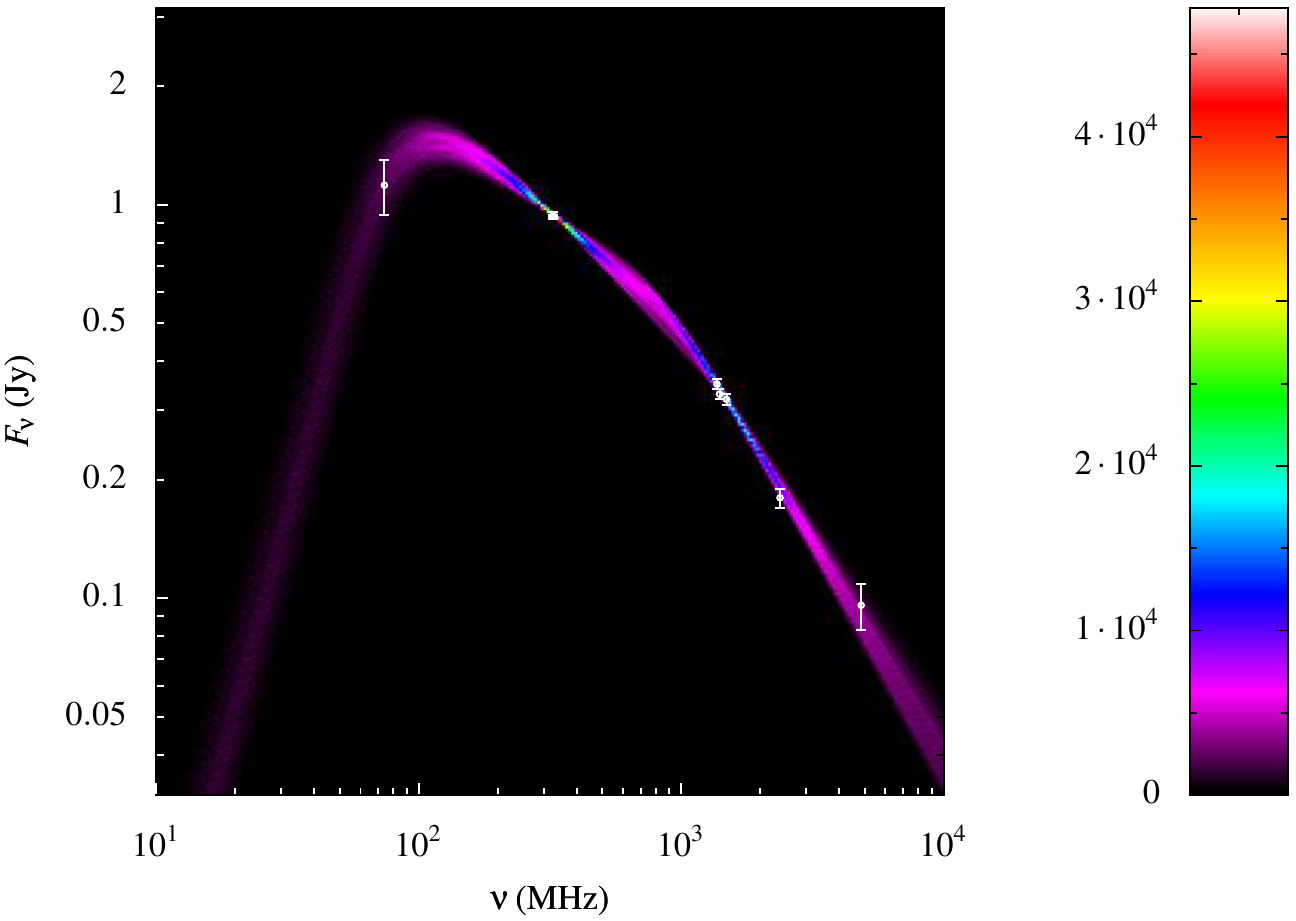}
\end{tabular}
\caption{NGC 7331}
\label{fig:ngc7331-fan}
\end{figure*}

The approach of plotting the model with maximum-likelihood parameters
however fails to capture the variation away from the
maximum-likelihood solution that is the main reason for the Bayesian
analysis approach in the first place. An alternative is to compute the
probability distribution of the flux density as a function of
frequency:
\begin{align}
  p(F_\nu| H,D)= \int {\rm d}\theta \cdot  F(\nu; \theta, H) p(\theta| H,D)
\end{align}
where as before $H$ is the hypothesis, i.e., the model for emission
and the priors on the model parameters; $p(\theta| H,D)$ is the
posterior distribution which is an output of the nested sampling; and
the integration over $\theta$ is of course over all its
dimensions. This probability distribution can then be plotted by
assigning frequency to horizontal position, flux density to vertical
position and the probability to colour-scale on the plot. Such
diagrams are sometimes called fan-diagrams in economics, and I adopt
that terminology here. This visualisation approach is shown in
Figures~\ref{fig:ngc628-fan}--\ref{fig:ngc7331-fan}.

\section{Implementation}

\begin{figure*}
\begin{tabular}{cc}
  Overall flux & Spectral index (at frequencies below the break) \\
  \includegraphics[clip,width=0.49\linewidth]{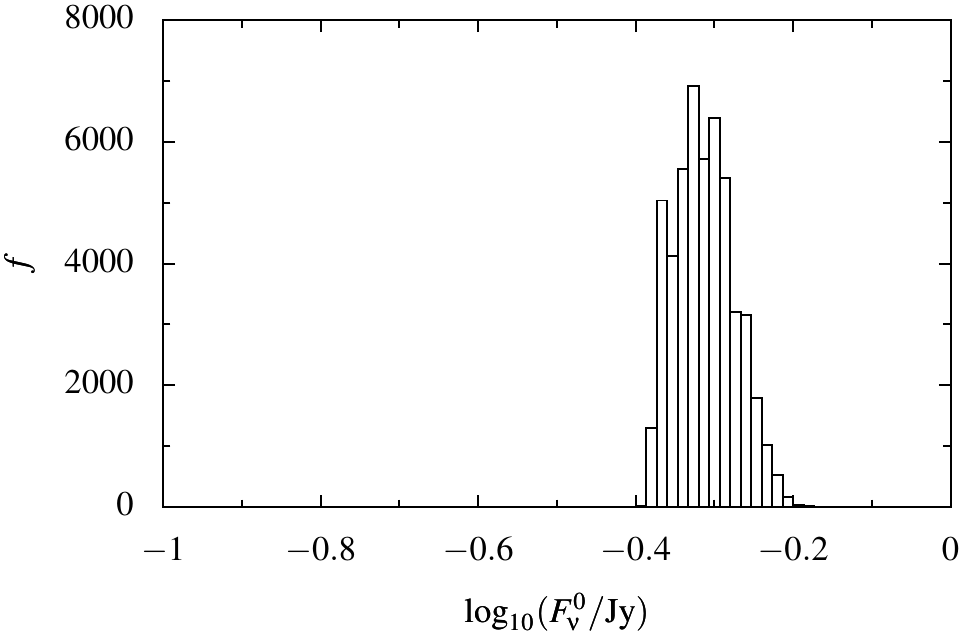}
  &
  \includegraphics[clip,width=0.49\linewidth]{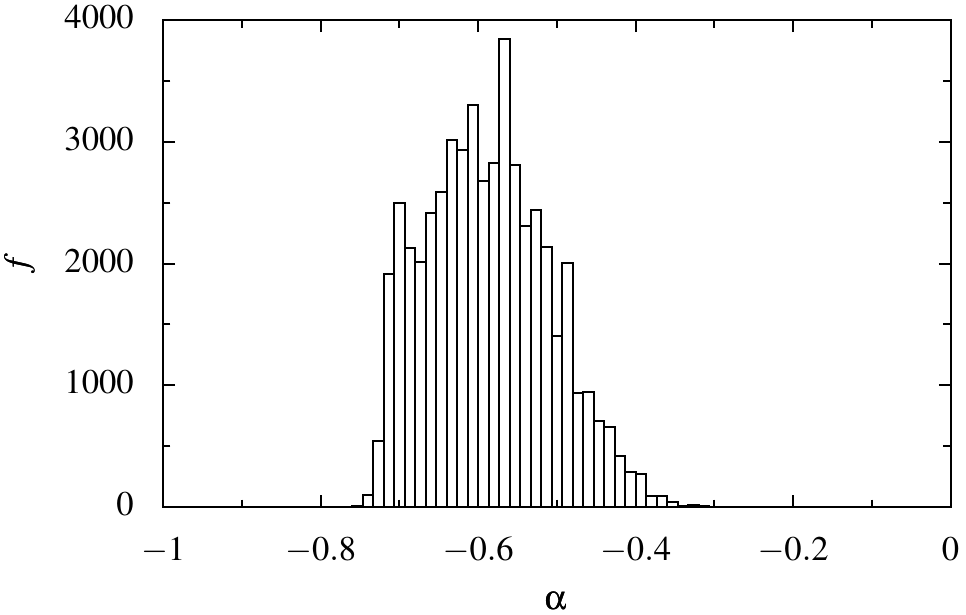}\\
  Frequency of the spectral break & Frequency at which self-absorption
  becomes important\\
  \includegraphics[clip,width=0.49\linewidth]{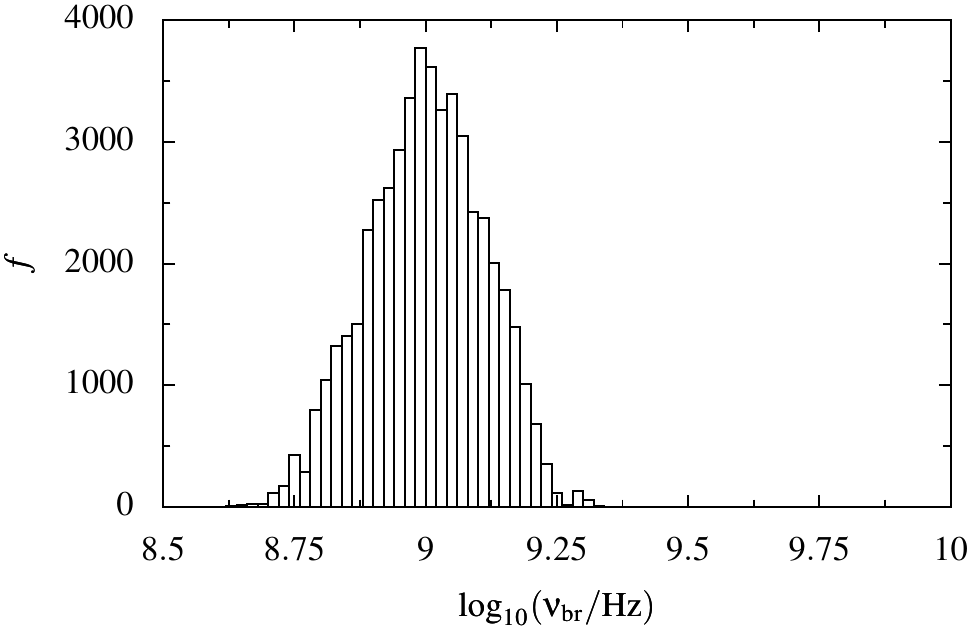}&
  \includegraphics[clip,width=0.49\linewidth]{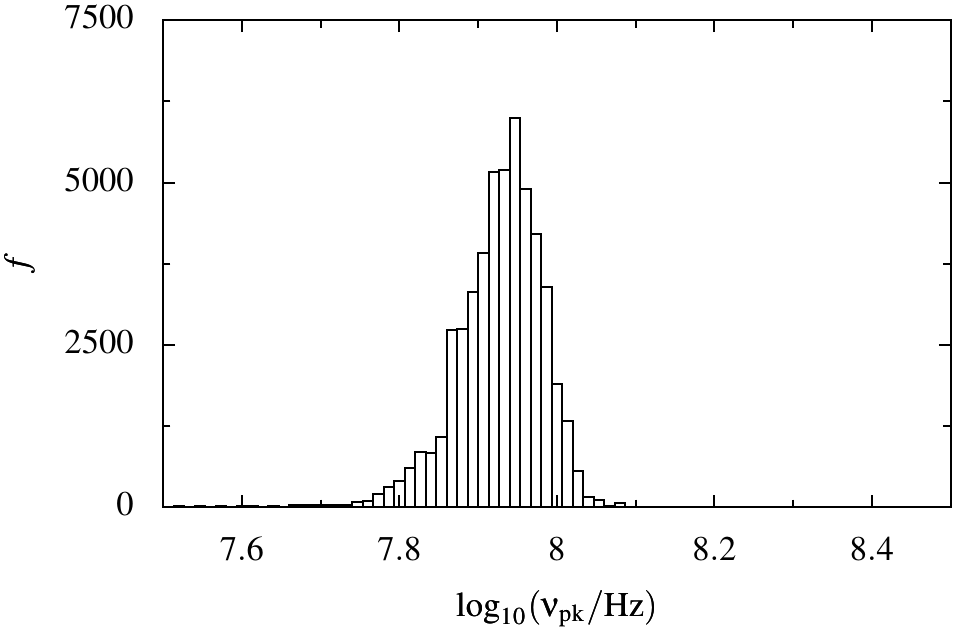}
\end{tabular}
\caption{Marginalised distributions for all of the parameters for the
  CI + SSA model of the radio spectrum of NGC 7331.}
\label{fig:ngc7331-margin}
\end{figure*}

The major computational parts of the algorithms described here are
implemented in the C++ programming language in two separate libraries:
\begin{itemize}
\item {\tt bnmin1} This is a minimisation and inference library that
  contains the nested-sampling algorithms and the supporting functions
\item {\tt radiospec} This is the library specialised for this
  application, and contains the models of radio spectra and
  descriptions of observations and their errors
\end{itemize}
The first of these, {\tt bnmin1}, is a general purpose
minimisation/inference library that may be used in a variety of
application. So far I have used this library for phase retrieval
holography (\citealt{2007A&A...465..679N},
\citealt{2007A&A...465..685N}), which is the application for which I
first started to develop the library; and for phase correction
algorithms for ALMA (\citealt{2009arXiv0903.1179N} and
\citealt{2009arXiv0907.4472N}). The library has been available to the
public under the GNU General Public License for a number of years and
although it has been downloaded occasionally I am not aware of other
public work using it. The majority of the functionality described in
this paper has only just been added in the release of the library that
accompanies this paper. The remaining functionality available in the
library includes Levenberg-Marquardt fitting and Markov Chain Monte
Carlo (MCMC).

The C++ parts of both of the libraries are relatively self-contained
with only two external dependencies: the Boost C++ libraries
\citep{BoostLibs}, and the GNU Scientific Library \citep{GSL}. The
build system of {\tt bnmin1} is based on the standard AutoTools chain,
while {\tt radiospec} is built using the SCons system.

The top level commands, such as which algorithms to use, to enter the
observed data, to control adjustable parameters etc are implemented in
the Python programming language. The interface between the C++
libraries and Python is generated automatically using the standard
SWIG\footnote{\url{http://www.swig.org/}} package described by
\cite{Beazley2003}. This architecture allows easy \emph{interactive\/}
use of the library. The supporting Python script for this application
is available as part of {\tt radiospec}.

All of the code is available for public download under the GNU General
Public License at
\url{http://www.mrao.cam.ac.uk/~bn204/galevol/speca/index.html}. I
should however point out some caveats:
\begin{itemize}
\item The packages need to be compiled from source and this typically
  requires some experience. The instructions posted on the above
  web-pages will be updated over time to explain how to tackle common
  problems with compilation that are reported to me
\item In order to analyse new observations using existing models, you
  will need to program in Python. There is no graphical or command
  shell user interface to this software
\item In order to create new models, you will need to program in C++
\end{itemize}
Included with the code is a script which reproduces the illustrative
examples given later in this paper.

\subsection{Implementation of radio spectra models}

The models of radio spectra are implemented as a polymorphic class
hierarchy in C++ in the {\tt radiospec} package. The base class
defining the interface is {\tt RadioModel} which in turn inherits from
the {\tt Model} class from the BNMin1 package. There are however only
two relevant ``virtual'' functions in the interface:
\begin{enumerate}
  \item {\tt double fnu(double nu) const} which computes the flux density at
    specified frequency {\tt nu}
  \item {\tt void AddParams(std::vector< Minim::DParamCtr > \&pars)}
    which defines the parameters of the model by adding them to the
    vector of parameter definitions {\tt pars}
\end{enumerate}
Any new models with which {\tt radiospec} is extend must properly
define these two functions to be useful.

The reason for adopting the polymorphic inheritance approach rather
than the more run-time efficient template approach is that the
polymorphic inheritance can be easily accessed and manipulated from
Python, allowing for example dynamic composition of several models
into a new, more complex, single model.

\subsection{Implementation of the likelihood function}

The likelihood function for the examples shown here is implemented in
the {\tt NormalLkl} class. This function combines an user specified
model (as a pointer to a type {\tt RadioModel} object), an user
specified set of observations (as an object of {\tt RadioObs} class)
and the usual Gaussian probability formula to compute the likelihood.

If a non-Gaussian likelihood function is required, it should be
implemented in a similar way to {\tt NormalLkl} class but it should
\emph{not\/} be derived from it. 

\subsection{Implementation of the priors}

In the current version of the code, only \emph{flat\/} priors which
are separable functions of single model parameter are
supported. Additionally, every model parameter must have \emph{some\/}
prior defined for it because this information is used to initialise
the nested sampling algorithm. Consequently, the user effectively has
to define a parameter-space prior `box' for each problem. 

The prior `box' is specified in the Python layer as a dictionary of
parameter names that map to a tuple specifying the lower and upper
bound of the parameter. An example is given in the file {\tt
  methodex.py}.

\subsection{Plotting}

The plotting of output is implemented using the PyX framework for
Python (\citealt{PyXWeb}). This library is able to directly write
PostScript (PS) and Portable Document Format (PDF) files and to run
LaTeX to generate properly type-set labels for the graphs. The main
benefits of this library for this application are that it can be used
directly from Python and that the output is of high-quality both
visually and in terms of the efficiency and readability of the output
PostScript code.

The routines that build on top of PyX to make the plots shown in this
paper are largely contained in the package {\tt PyHLP} which is
distributed separately from the other packages and can be downloaded
at \url{http://www.mrao.cam.ac.uk/~bn204/technotes/pyhlp.html}.

\section{Results}

The data I used to illustrate this approach are taken from
\cite{2009A&A...503..747P}.  They are global spectra of three
late-type galaxies with measurements at 5 to 7 frequencies in the
range 50\,MHz--5\,GHz. I have taken the measurements and measurement
errors listed by \cite{2009A&A...503..747P} without any further edits
and without referring to the original sources for the archival data
that they used.

The spectra of all three galaxies were processed as described above,
including the use of the priors listed in Equations
\ref{eq:priors-bg}--\ref{eq:priors-end}. The results are presented as
the fan-plot for each model, together with the evidence value in
Figures~\ref{fig:ngc628-fan}--~\ref{fig:ngc7331-fan}. For NGC 7331, in
Figure~\ref{fig:ngc7331-margin} I also plotted the marginalised
distributions of the model parameters.

\subsection{NGC 628}

The radio spectrum of the galaxy is relatively featureless and can be
by-eye seen to be reasonably close to a pure power-law. The analysis
presented here also reaches this expected conclusion.

The fan-diagrams of all four models for this object are shown in
Figure~\ref{fig:ngc628-fan}.  The combination of model and prior with
the highest evidence value is the power-law case, shown in the
upper-left panel of this figure. This means that this simple model is
the best model for the true underlying process given the observations
out of the four models considered here.

The continuous injection model fan-diagram in the lower-left panel
shows that it can explain the data in two ways: either the break is at
a frequency higher than the highest observations, or, the break is at
around 300\,MHz and the lowest-frequency data point is predicted with
a significant error. Therefore there in principle remains a
possibility that this galaxy has a highly aged electron population
with an intrinsic injection index of about $\alpha \sim -0.2$. This is
however unlikely given the lower evidence value of this model compared
to the power law.

Finally it can be seen that the absorbed models on the right hand side
of Figure~\ref{fig:ngc628-fan}, i.e., the upper-right and lower-right
plots, do not describe the data as well as the non-absorbed
models. This is part because the minimum frequency of the turnover was
set at 30\,MHz by priors.

\subsection{NGC 3627}

The fan-diagrams for this galaxy are shown in
Figure~\ref{fig:ngc3627-fan}.  The model with the highest evidence is
the continuous-injection model without absorption at low
frequencies. The lower evidence of the CI+SSA model indicates that
with these data, there is no evidence for absorption in this source.

\subsection{NGC 7331}

The fan-diagrams for this galaxy are shown in
Figure~\ref{fig:ngc7331-fan}. What is noticeable for this galaxy is
that the most complex model, the continuous-injection with synchrotron
self-absorption model (CI+SSA), has the highest evidence value by
several orders of magnitude. This high evidence value implies that
this complex model must be preferred given the available data. The
fan-diagram of the CI+SSA model, in the lower-right panel of the
figure, shows that it reproduces well the observed features of the
spectrum while all of the others fail to reproduce one or more
features.

The marginalised distributions of the parameters of the CI+SSA model
are shown in Figure~\ref{fig:ngc7331-margin}. It shows that the
distributions of all of the parameters are well constrained, although
at least the spectral index and the break frequency show non-Gaussian
distributions. This should be interpreted to mean that care must be
taken when using a simple single-value error estimate for these
parameters in further calculations.

\section{Summary}

Fitting of radio spectra of galaxies is a topic that is
computationally relatively simple, since most models are either
analytic or contain simple one-dimensional integrals. A proper
statistical analysis is however not entirely straightforward for a
combination of reasons:
\begin{enumerate}
\item There are few measurement points (typically 3--10)
\item Errors are often non-Gaussian (e.g., because they are dominated
  by calibration errors) and are sometimes not well quantified
\item There are many different models that could be tried
\end{enumerate}

An attractive way to tackle this problem is using Bayesian analysis
because it provides:
\begin{enumerate}
  \item A rigorous theoretical framework
  \item Objective model selection
  \item A natural way to introduce physical constraints on model
    parameters through priors
  \item Full probability distributions for each model parameter
  \item A complete picture of any degeneracies in the model parameters
\end{enumerate}

In this paper I have described a publicly available computer code
which implements such Bayesian analysis of spectra using the nested
sampling algorithm developed by \cite{Skilling2006}.  This algorithm
allows efficient calculation of all of the outputs of Bayesian
analysis including the evidence value and the full joint distribution
of all parameters. This means analysis is computed quickly and without
the need to guide it `by hand', by for example carefully choosing
starting positions.

The described code also allows visualisation of how well the each
model explains the data using fan-diagrams. I believe this little-used
approach to visualisation allows a good and intuitive understanding of
implications of a particular distribution of model parameters.

The code described is already being used for several of projects in
radio astronomy but I expect it could useful for quite a broad range
of applications. Such application to new areas will no doubt lead to
discoveries of errors and shortcoming in the code and I would very
much appreciate to be notified these at
\url{mailto:b.nikolic@mrao.cam.ac.uk}. If you obtain useful results
from the code without finding any errors or shortcomings, than of
course I would be even more happy to hear from you, at the same email
address, and can place a link to your paper on the web-pages.

\bibliographystyle{mn2eurl} 
\bibliography{90ghzgals.bib}

\section*{Publication information}

As an experiment, I will be publishing this paper on arXiv only. This
is in part because future revision of this paper is likely to be
necessary, once the code is used more extensively both by us and
hopefully by the general community.

In lieu of the normal referring process, I would be happy to hear from
readers on any aspect of the paper and incorporate all corrections and
(at least constructive) suggestions in any future versions. All of
these will be credited unless requested otherwise. Alternatively if
you have more extensive comments I suggest you use the arXiv trackback
mechanism.

\label{lastpage}
\end{document}

%% file: versioninfo.tex
\newcommand{\bzrdate}{2009-12-11 19:02:01 +0000}
\newcommand{\bzrrevno}{47}